# Final Size Planar Edgeless Silicon Detectors for the TOTEM Experiment

E. Noschis,[a,f*] E. Alagoz,[a] G. Anelli,[a] V. Avati,[a,d] V. Berardi,[g] V. Boccone,[h] M. Bozzo,[h] E. Brucken,[f] A. Buzzo,[h] M.G. Catanesi,[g] R. Cereseto,[h] S. Cuneo,[h] C. Da Viá,[c] M. Deile,[a] R. Dinapoli,[a] K. Eggert,[a] N. Egorov,[j] I. Eremin,[i] F. Ferro,[h] J. Hasi,[c] F. Haug,[a] J. Heino,[f] P. Jarron,[a] J. Kalliopuska,[f] J. Kašpar,[b] A. Kok,[c] Y. Kozlov,[j] W. Kundrát,[b] K. Kurvinen,[f] R. Lauhakangas,[f] M. Lokajíček,[b] T. Luntama,[f] D. Macina,[a] M. Macrí,[h] S. Minutoli,[h] L. Mirabito,[a] A. Morelli,[h] P. Musico,[h] M. Negri,[h] H. Niewiadomski,[a,c] F. Oljemark,[f] R. Orava,[f] M. Oriunno,[a] K. Österberg,[f] A.-L. Perrot,[a] R. Puppo,[h] E. Radermacher,[a] E. Radicioni,[g] H. Saarikko,[f] A. Santroni,[h] G. Sette,[h] A. Sidorov,[j] P. Siegrist,[a] J. Smotlacha,[b] W. Snoeys,[a] C. Taylor,[d] S. Watts,[c] J. Whitmore[e]

[a]CERN, CH-1211 Geneva 23, Switzerland

[b]AVČR, Institute of Physics, CS-18040 Prague 8, Czech Republic

[c]Brunel University, Uxbridge, Middlesex UB8 3PH, UK

[d]Dept. of Physics, Case Western Reserve University, Cleveland,OH 44106, USA

[e]Dept. of Physics,Penn State University, University Park, PA 16802, USA

[f]Helsinki Institute of Physics and Department of Physical Sciences,P.O. Box 64, FIN-00014 University of Helsinki, Finland

[g]INFN Sez. di Bari and Politecnico di Bari, Via Orabona 4,I-70126 Bari, Italy

[h]INFN Sez. di Genova and Universita di Genova, Via Dodecaneso 33, IT-16146 Genoa, Italy

[i]Ioffe Physico-Technical Institute, Polytechnicheskaya Str. 26, 194021 St-Petersburg, Russia

[j]Russian Institute of Material Science and Technology,124460 Moscow, Zelenograd, Russia

**Abstract**

The TOTEM experiment will detect leading protons scattered in angles of microradians from the interaction point at the Large Hadron Collider. This will be achieved using detectors with a minimized dead area at the edge. The collaboration has developed an innovative structure at the detector edge reducing the conventional dead width to less than 100 μm, still using standard planar fabrication technology. In this new development, the current of the surface is decoupled from the sensitive volume current within a few tens of micrometers. The basic working principle is explained in this paper. Final size detectors have been produced using this approach. The current-voltage and current-temperature characteristics of the detectors were studied and the detectors were successfully tested in a coasting beam experiment.

———

* Corresponding author. Tel.: +41-22-76-788-42; fax: +41-22-76-709-70; e-mail: elias.noschis@cern.ch.

## 1. Introduction

The TOTEM experiment [1],[2] will detect leading protons scattered in angles of microradians at the Large Hadron Collider (LHC). This will be achieved using 'edgeless' microstrip tracking detectors with a minimized dead width located in special insertions called Roman Pots [3]. The detectors will approach the beam center to a distance of 10 σ + 0.5 mm (=1.3 mm), and the dead space near the detector edge must not exceed 100 μm to ensure the required acceptance in leading proton detection.

The working principle of planar edgeless detectors has been introduced in [1]. The typical distance between the strips of the detector and the cut edge is 50 μm, allowing the sensitive volume to be as close as 60 μm from the cut edge of the detector [4].

Final size planar edgeless detectors for the TOTEM experiment have been produced and tested. Their electrical and thermal performance has been studied. The detectors were inserted in a Roman Pot prototype attached to the beam pipe of the Super Proton Synchrotron, and their response to high energy protons in a coasting beam experiment was investigated.

## 2. Planar edgeless detector: working principle

Standard planar silicon detectors have a typical insensitive border region around the sensitive area of 0.5 mm occupied by guardrings. This voltage terminating structure drops the voltage gradually between the detector sensitive area and the detector edge.

The basic idea of this new approach is to reduce the insensitive border below 100 μm by applying a large fraction of the detector bias across the detector chip cut through an outer current terminating ring (CTR) that collects the major part of the resulting surface generated current $I_{CTR}$ (see Figure 1).

A ring placed between the CTR and the strips, called clean-up ring (CR) is biased at the same potential as the CTR. The leakage current $I_{CR}$ collected at the CR consists mainly of a small fraction of the current generated at the cut surface due to diffusion mechanism. This structure decouples the high leakage current generated at the detector cut edge from the sensitive volume. This is evidenced by measurements presented in Section 3.1.

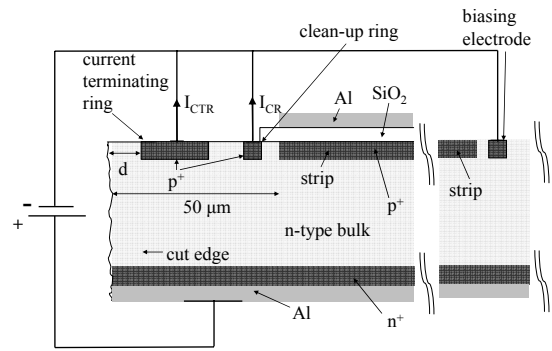

Figure 1: Cross-section of a silicon detector with a current terminating structure in the plane parallel to the strips and its biasing scheme.

The strips are biased by means of punch-through mechanism [5] between the biasing electrode and the strips. One relies on the fact that two nearby diodes will only withstand a difference of a few volts before a current starts flowing between them through the bulk. The biasing ring collects the bulk generated current from the sensitive area. In contrast with standard guard ring structures which provide voltage termination, this structure terminates the current, and therefore we have called it "Current Terminating Structure" (CTS). This structure has been tested on small size devices with $d$ = 0 (see Figure 1) and has shown to work successfully [4].

## 3. Final size planar edgeless detector

Final size planar edgeless detectors have been developed and produced in a joint effort of the TOTEM group at CERN and the Megaimpulse, a spin-off company from the Ioffe Physico-Technical Institute in St. Petersburg (RUSSIA). These devices are microstrip detectors with 512 strips, with a pitch of 66 μm processed on very high resistivity *n*-type silicon wafer (>10 kΩ·cm), 300 μm thick. All of them have a CTR as described in Section 2 on one edge (the bottom edge in Figure 2) and AC coupled strips biased through the biasing electrode. In these devices, the biasing electrode is electrically connected to the clean-up ring. For these detectors, the nominal *d* value was set to 20 μm. A picture of a final size detector mounted on the readout board is shown in Figure 2.

The strips (sketched in Figure 5) are at a 45° angle from the vertical axis. Flipping around the vertical axis yields strips reoriented orthogonally.

### 3.1. Electric Characterisation

The produced samples were tested on a sample holder with CR and CTR both bonded to allow current/voltage (I-V) and current/temperature (I-T) measurements using the biasing scheme of Figure 1. Typical values of currents measured at the CTR and the CR, $I_{CTR}$ and $I_{CR}$ in Figure 1, are shown in Figure 3.

There is a difference of four orders of magnitude between the current at the CR and the CTR. This is strong evidence that a very large fraction of the leakage current generated at the surface is collected by the CTR, preserving the detector active area from almost any surface current, which would make the detector operation impossible.

The low current flowing in the CR confirms the validity of the current termination approach: the sensitive bulk, even if it extends to a few tens of micrometers from the cut edge is free of the large current flowing at the surface.

These detectors deplete fully at the reverse bias of ~30 V and have shown to be stable for biases higher than 150 V. In order to study the nature of the bulk and the surface currents, their behaviour with different temperatures was also measured. A typical temperature dependence of these two currents is shown in Figure 4 for a reverse bias of 50 V.

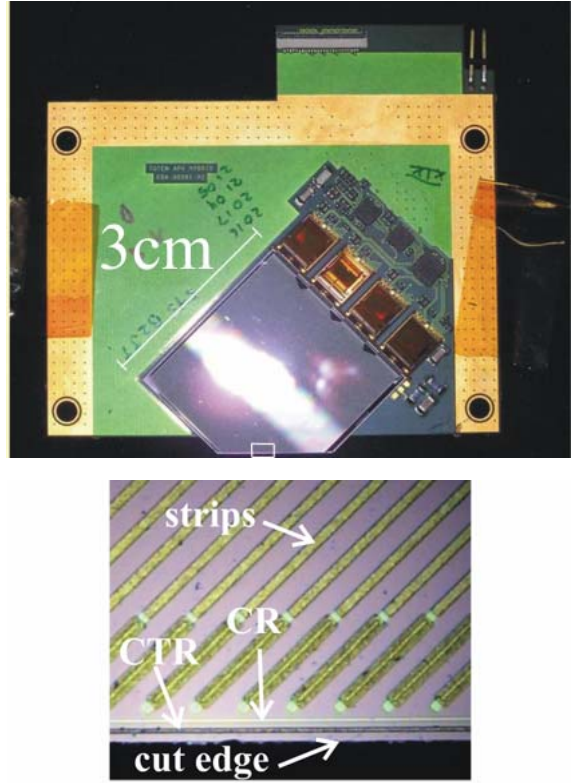

Figure 2: (Top) Picture of the final size planar edgeless detector and the readout board. Only the bottom edge has a CTS. The strips are oriented at a 45° angle from the vertical axis. (Bottom) Enlarged picture of the rectangular box in the top figure. The characteristic features of the CTS are clearly visible.

The current collected by the CR shows an exponential increase with the temperature. The current flowing through the CTR increases with temperature but at a lower rate than the bulk current. The average activation energy of 0.6 eV gathered from $I_{CR}(T)$ in Figure 4 allows to conclude that the current measured at the clean-up ring consists in mainly bulk generated current. These data still confirm the basic idea of the CTS and the decoupling of the bulk current from the surface current (a difference of three orders of magnitude even higher at lower temperatures) and show that a further reduction of the surface current with temperature is possible.

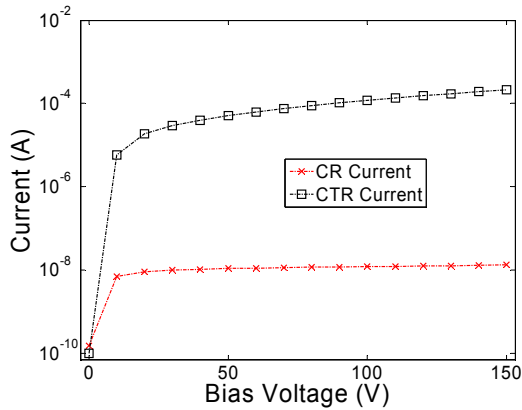

Figure 3: Current vs. voltage characteristics measured at the current terminating ring and the biasing ring. The measurement was done at -12°C.

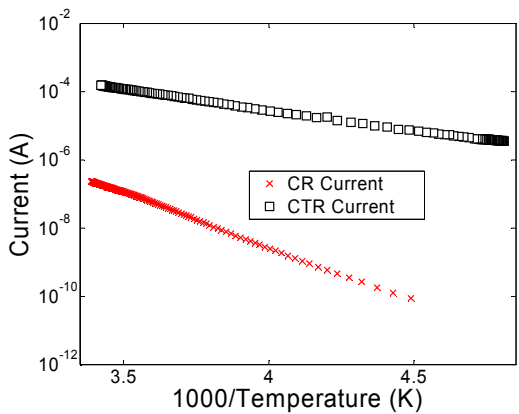

Figure 4: Arrhenius plot for the current flowing through the biasing ring and the current terminating ring for a reverse bias of 50 V.

## 4. Beam Tests

### 4.1. Experimental Setup

The edgeless planar silicon detectors were studied in a coasting beam experiment in the Super Proton Synchrotron (SPS) at CERN. A Roman Pot unit prototype consisting of a vacuum chamber equipped with two vertical insertions, top and bottom pots, located along the beamline of the SPS accelerator ring hosted a set of detectors each. The secondary vacuum of the detectors was separated from the primary beam vacuum by a thin window foil (140 μm for the bottom pot and 210 μm for the top one).

The vertical position of the Roman Pots was adjusted by micrometric screws controlled by stepping motors allowing one set of detectors to approach the beam axis from the top, and the other one from the bottom. Both sets of detectors consisted of four pairs of edgeless detectors, each pair consisting of detectors mounted together back-to-back, defining an (x,y) coordinate point. Three pairs were used for tracking and were read out with analogue APV25 chips [6]. One pair was read out with digital VFAT chips delivering the fast-or signal of all 512 strips [7]. The latter were used for triggering the data acquisition system in coincidence with the sum signal of the four pick-up electrodes of a beam position monitor located close to the detectors.

Three different bunch structures were tested in the SPS accelerator: 1 single bunch in the accelerator ring, 4 bunches equally spaced, and 4 equally spaced trains of 4 bunches of $8 \times 10^{10}$ 270 GeV protons with a revolution period of 23 μs [8].

### 4.2. Analysis and Results

Detector data were taken with the two pots moving independently between 6 mm and 14 mm ($\sigma_{beam}$ = 0.8 mm) from the beam pipe centre. Beam halo protons were detected at typical rates of 3 kHz. Figure 5 shows the halo profiles measured by two orthogonal detector planes of this pot.

The detectors had a typical signal-to-noise ratio of 22. There were a few noisy channels on both detectors which have been removed from the profiles.

## 5. Conclusion

Final size edgeless planar detectors with a current terminating structure for the TOTEM experiment at the LHC have been produced. Their thermo-electrical properties have been studied. The detectors were tested successfully in a coasting beam experiment under conditions similar to the ones of the LHC. Confrontation of beam test results with metrology measurements to determine the detector efficiency from the edge is on progress and will be the subject of a future publication.

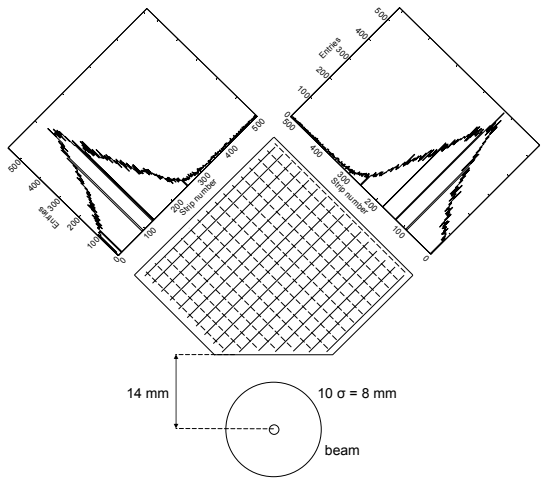

Figure 5: Profile of the beam halo as seen by two orthogonal detector planes at a distance of 14 mm from the beam center. The data were taken with the bottom pot and the picture has been rotated by 180° around the beam axis for more convenience.

## Acknowledgments

We would like to thank V. Eremin for his great support in the detector development. Also, the authors are thankful to I. McGill, R. de Oliveira and T. Souissi for their valuable technical contributions.